# An XML Driven Graphical User Interface and Application Management Toolkit

Matthias Clausen (DESY & SLAC), Piotr Bartkiewicz (DESY & FPNT/AGH), Alexei Dmitrovski, Albert Kagarmanov (DESY & IHEP), Zoltan Kakucs (DESY), Greg White, Hamid Shoaee (SLAC)

Abstract

In the past, the features of a user interface were limited by those available in the existing graphical widgets it used. Now, improvements in processor speed have fostered the emergence of interpreted languages, in which the appropriate method to render a given data object can be loaded at runtime. XML can be used to precisely describe the association of data types with their graphical handling (beans), and Java provides an especially rich environment for programming the graphics. We present a graphical user interface builder based on Java Beans and XML, in which the graphical screens are described textually (in files or a database) in terms of their screen components. Each component may be a simple text read back, or a complex plot. The programming model provides for dynamic data pertaining to a component to be forwarded synchronously or asynchronously, to the appropriate handler, which may be a built-in method, or a complex applet. This work was initially motivated by the need to move the legacy VMS display interface of the SLAC Control Program to another platform while preserving all of its existing functionality. However the model allows us a powerful and generic system for adding new kinds of graphics, such as Matlab, data sources, such as EPICS, middleware, such as AIDA[1], and transport, such as XML and SOAP. The system will also include a management console, which will be able to report on the present usage of the system, for instance who is running it where and connected to which channels.

## 1 MOTIVATION

The work on the JoiMint (Java Operator Interface and Management Integration Tool) application started at SLAC as a proof of principle study for the COSMIC[2] project. Special emphasis was given to keeping the functionality of the existing applications. An important aspect of these applications is the communication between the operator interface and the application on the mainframe itself. In this case the operator can configure the settings for the display. According to these settings the database will be queried to extract all necessary information and to retrieve the live data from the control system as well as all information to format and render the graphics.

In a distributed computer environment the situation is more complex because all the communication between the client and the application preparing the graphic has to run over a protocol. Even more important is the necessary 'push' model for the graphics which has to be defined by the application while the rendering process is performed by the client tool.

This was the starting point for the study of a graphical toolkit which will clearly separate the definition of a display from the rendering process and the data access from the graphic update. In addition, the study should show whether the management of client applications can be integrated into such a toolkit.

## 2 DEVELOPMENT ENVIRONMENT

Since the toolkit should in the end also run as an Applet, the decision to use Java as the basic language for the development was obvious. Visual Café was chosen as the development environment because of its file oriented environment, which helps to coordinate development efforts in a widely distributed project. While the work was started at SLAC, it was finished at DESY as a joint project between several groups at DESY and the controls group at SLAC.

## 3 BASIC FUNCTIONALITY

### 3.1 Operation Mode(s)

JoiMint is only running in one operation mode – the 'run' mode. Any changes to the set of active objects can be modified on the fly. Configurations can be loaded from disk or over the network, new synoptic windows can be opened and new objects can be placed on their plane. They can be dynamically connected to or removed from one or more underlying control systems. Actual configurations can be saved to disk. Before saving, configurations can be edited in an edit window.

## 3.2 Graphic Objects

Especially in the Java environment graphic objects are seen as Java beans only. Typically the object oriented capability of these beans is extended to support not only scalar values coming from the underlying control system but also complex device information. One of these implementations is known as Abeans[3]. The advantage of this approach is the close relation to the control system. On the other hand it implies a strong binding between graphic and controls properties. The JoiMint approach tries to keep this binding as loose as possible. Each property of the graphic objects can be connected to individual properties of one or several underlying control systems. This implies the following states for each property:

- **Static**
  The property has a fixed (static) value. It can be only changed from the 'edit property' window.
- **Dynamic**
  The property is dependent on the dynamic behavior of a device. Updates will be sent to the object on scan or asynchronously.
- **Active**
  The property is controlling a dynamic device. The remote device as well as any properties of other objects registered with this device name will receive updates.

Objects in general have no direct dependency on the core code. They have to implement an interface providing a certain set of mandatory methods like loadConfig() and saveConfig().

Since objects register themselves with their object and (propertyUpdate)method(s) in the central registry, any dynamic properties of any object can be accessed for data updates.

Following these loose implementation rules any existing object can be easily integrated into JoiMint. Future versions will also support the dynamic loading of new objects with seamless integration.

## 3.3 The Central Registry

The registry is the 'heart' of JoiMint. Each object is registered in the registry 'as an object'. This way any object can be accessed 'by name' through the registry. Commands can be configured and sent between objects on the fly. To uniquely identify objects by their name, the name of graphic objects on a window plane consist of 'theWindowsName|theObjectsName'. In addition to this registration, properties of dynamic objects are registered with their associated device name. This way objects are completely decoupled from the data source. Whenever an updated device name/value pair is sent to the registry, it will locate the device name and send the new value to all objects which are registered with this device. This way each device (or record-) name will only be registered once, while the updates can be sent to many objects.

## 3.4 Data Sources

By definition, JoiMint supports the http protocol. Thus http servers are native data sources for JoiMint. A special http server has been developed to support dynamic data updates as well as to access archived data. The initial implementation supports the EPICS channel access protocol and reads archived data from the EPICS channel archiver. The next version will also support monitors and additional protocols like TINE and DOOCS that are used at DESY.

In addition local data sources can be defined. One such local source is a random data generator, which can be used to test and demonstrate the behavior of objects. Also objects can be the source of dynamic data, like the planned calculation object which will compute a dynamically defined formula depending on up to ten dynamic input values. This functionality is also known as the 'calculation record' in EPICS.

Furthermore the direct connection to control systems' protocols can be implemented if JoiMint runs in the application mode. Setups with the channel access protocol and a CORBA communication were tested successfully.

Http based data sources can be added on the fly. Since the internal naming structure consists of 'dataSource|deviceName', the following name would be a well formed JoiMint device name: "http://JoiMint.desy.de:8081|deviceName". In practice http-host:port addresses would be used in conjunction with alias names like 'EPICS' or 'ARCHIVER'. This way, a stored configuration can be used remotely (EPICS:== http-host:port) or locally (EPICS:==channel access) without modification.

## 3.5 XML

XML is used in several places in JoiMint. All configuration files are written in XML. This includes the configuration of JoiMint as a tool as well as the configuration of synoptic displays. Internal SAX parsers are used to read these files. While the main application is reading and writing its configuration as a whole, the configuration of each graphic object is parsed by the individual object itself. This way neither JoiMint nor the parent window has any knowledge of the objects specific properties or their XML representation. New functionality can be added to objects without changing the configuration. Just enabling new features and saving the new property configuration will be sufficient. Of course completely new objects can be easily integrated this way also.

XML is additionally used as a data representation layer for http communication. The overhead of ASCII communication is more than compensated by the transparent access across computer platforms. XML over http was also chosen because it is (so far) not used as a medium to transport viruses. Firewalls are used to filter traffic in many sites. Releasing additional http ports for external access seems to be more appropriate for this purpose than unknown – and maybe unsafe – binary protocols.

## 4 MANAGEMENT

The management of running applications is a problem on it's own. It is only possible to manage running applications if they are registered in a central place. The console manager at SLAC is providing this kind of service. A central registry can be queried to find out which application is running on which console. Misbehaving applications can be identified and shut down if necessary. These features initiated the idea to implement this kind of service into JoiMint as a basic functionality. JoiMint, providing a 'light' http server on its own can be remotely queried for the most important parameters, like: number of open connections, number of updated values/sec or number of concurrent threads currently running. This information will be sent as 'pure' http if a native http client like Netscape is connected. It will be sent as XML if a JoiMint client connects. In this case JoiMint will first send the configuration of the 'JoiMint Management Page' and then send the corresponding data which can also be queried by the calling JoiMint client.

### 4.1 Smart Devices

The trend to load a light http server on any smart device connected to the Ethernet is obvious in industry. Ethernet based PLC's are already equipped with http servers for diagnostic purposes. Separating diagnostic data from control data is a current trend. Smart power supplies with a tight connection to the control system and an additional http/XML connection to, for instance, JoiMint, seems to be a promising view into the future.

### 4.2 Global Accelerator Networks [4],[5]

The field of http/XML based applications is not only limited to smart devices. The other extreme is complete accelerators being built by international collaborations. State of the art tools for remote diagnostics and controls will be necessary in many areas. JoiMint might be an example of such a tool.

## 5 FUTURE

JoiMint has been designed to be open for future joint developments. The open architecture of the data source as well as the object interface will ease the integration of new protocols and the implementation of new objects. The set of graphic objects will be increased and enriched with additional functionality.

Possible candidates for complete applications which could be integrated are for instance Java Mathematica clients or light versions of EPICS tools like alarm handler or the integration of a CMLOG client.

As an additional feature of JoiMint the recording and playback of complete sequences of commands and data is foreseen for a future release.

## REFERENCES

[1] Robert Sass, et al. Aida - Accelerator Integrated Data Access. An NLC Middleware Concept. ICALEPCS 2001 San Jose CA, Proceedings

[2] M. Clausen et al., COSMIC - The SLAC Control System Migration Challenge. ICALEPCS 2001 San Jose CA, Proceedings

[3] Phil Duval, New abeans for TINE Java Control Applications. ICALEPCS 2001 San Jose CA, Proceedings

[4] R. Bacher, What are the Controls Requirements for the Global Accelerator Network. ICALEPCS 2001 San Jose CA, Proceedings

[5] F. Willeke, How to Commission, Operate and Maintain a Large Future Accelerator Complex from Far Remote. ICALEPCS 2001 San Jose CA, Proceedings